\documentstyle[twocolumn,aps,prl,epsf,floats]{revtex}
\begin{document} 
\title {
Theory of the leading edge gap in underdoped cuprates}
\author{Andrey V. Chubukov}
\address{
Department of Physics, University of Wisconsin, Madison, WI 53706}
\date{today}
\maketitle

\begin{abstract}
We present the theory of the leading edge gap in the normal state of underdoped
high-$T_c$ materials. The consideration is based on a
magnetic scenario for cuprates. We show that as doping decreases,
the increasing interaction with paramagnons  gives rise to a near
destruction of the Fermi liquid and this in turn yields precursors to $d-$wave 
pairing.
We argue that the leading edge gap at $\sim 30 meV$ and  a broad maximum
 in the spectral
function at $\sim150 meV$ are byproducts of the same physical phenomenon.
\end{abstract}

One of the most intriguing experimental facts about underdoped
cuprates is that they display superconducting properties
already at temperatures which can be
few times larger than $T_c$. This 
phenomenon, which has 
been observed in the NMR, transport and optical measurements~\cite{review},
is most  directly seen in photoemission
experiments on $2212Bi$ compounds: the leading edge
of the photoemission curve
remains at a finite distance from zero 
energy well above the actual $T_c$ and displays an angular dependence,
similar to that of a true $d-$wave superconducting gap~\cite{camp,shen}.
 This pseudogap behavior is however rather
peculiar as the leading edge gap (LEG) is not accompanied by the quasiparticle
peak.
Instead, the spectral function is rather flat above the 
LEG and only 
displays a broad maximum at a frequency $\sim 150 meV$ 
which is $5$ times larger than the 
gap. This last frequency
is comparable to the spin exchange integral $J$, and this
caused the speculations that the high-frequency maximum
can be due to the precursors to antiferromagnetism~
\cite{morr,laughlin}.
 A challenging observation for
this conjecture is that the LEG
 and the broad maximum at higher frequencies
seem to emerge at the same doping concentration~\cite{shen2} and 
therefore are likely to be byproducts of the same physical phenomenon.

In this paper, we show that both,  
the LEG and the broad maximum at higher frequencies
 can simultaneously be explained in the magnetic scenario for cuprates. 
This scenario implies that the low-energy physics of cuprates
is described by 
the spin-fermion model
in which itinerant fermions interact with their collective spin degrees of
freedom by 
\begin{equation}
{\cal H}_{s-f} = g 
\sum c^{\dagger}_{k,\alpha} {\vec \sigma}_{\alpha,\beta}
c_{k+q,\beta} {\vec S}_{-q}.
\label{1}
\end{equation}
Here $g$ is the coupling constant  which is assumed to increase
as the system approaches half-filling, and 
$\sigma_i$ are the Pauli matrices. 
This model can be obtained from the underlying Hubbard-type model by
integrating out high-energy fermions and performing the RPA summation
in the particle-hole channel.

The propagator for low-energy fermions is
assumed to have a Fermi-liquid form
$G_0 (k,\omega_m) = Z_0/(i\omega -
\epsilon_k)$ where for $\epsilon_k$ we use a tight binding form 
$\epsilon_k = -2t (\cos k_x + \cos k_y) - 4 t^{\prime} \cos k_x \cos k_y - \mu$ 
%with $t^{\prime} \sim -0.2t$ 
consistent 
with the photoemission experiments on the overdoped
cuprates~\cite{flatdisp}. 
The collective variables
 $S_q$ are characterized by their bare spin susceptibility
$\chi_0 (q, \omega) = {\tilde \chi}/(1 + ({\tilde q} \xi)^2 - 
\omega^2/\Delta^2)$, 
where $\xi$ is the spin correlation length, 
$\Delta = v_s \xi^{-1}$ where $v_s$ is the spin-wave velocity, 
and ${\tilde q} = Q -q$ where $Q$ is either equal or very close to the
antiferromagnetic momentum $(\pi,\pi)$.
Observe that we did not introduce the  damping term
$\propto i\omega$ into $\chi_0$. We argue that for $T\ll J$,  
the key source of spin damping 
is the decay of a spin fluctuation into a particle-hole pair.
In this situation, spin damping is not an input parameter in the theory, 
but rather should be
obtained self-consistently within the spin-fermion model. 

The quasiparticle residue $Z_0$
is generally a function of $T$, and it becomes a constant
only below some $T^*$~\cite{review}, when 
quantum fluctuations start to dominate over classical fluctuations.
In this paper, we assume for simplicity that 
classical fluctuations can be completely 
neglected below $T^*$ 
and set $Z_0(T) = Z_0(T^*)$. 
In practical terms, this implies that we in fact will be computing the LEG 
right above $T_c$ where it is maximal and will  not discuss how
this gap is destroyed by thermal fluctuations.
For the rest of the paper we
absorb both $Z_0 (T^*)$ and ${\tilde \chi}$ into the coupling constant: 
$g Z_0 (T^*) \sqrt{\tilde \chi} \rightarrow g$. 

We now proceed with the calculations. 
Our strategy is
the following: we first demonstrate that when the coupling exceeds 
some typical value $g_0$,                
the self-energy corrections
to the fermionic propagator nearly completely destroy the Fermi liquid in the
vicinity of $(0,\pi)$ and related points.
 Then we use the renormalized form of $G$ to compute the pairing 
susceptibility in the
$d_{x^2 -y^2}$ channel. We show that this susceptibility is attractive, 
and for $g > g_0$
yields a $d-$wave LEG above $T_c$. 
Finally, we show how  the leading
edge gap transforms into a true superconducting gap below $T_c$.

We begin by reviewing  the earlier results for
the fermionic and bosonic self-energies in the spin-fermion 
model~\cite{andrey,lev}.  The bosonic 
self-energy gives rise to a  damping term in 
 the full spin susceptibility: 
$\chi^{-1} (q,\omega) = \chi^{-1}_0 (q,\omega) + 
i {\tilde \chi}^{-1}~\omega/\omega_{sf}$ where 
%for $t^{\prime} \ll t$,
$\omega_{sf} \approx (3/16) v \xi^{-1} (g_0/g)^2$,  
$g^2_0 = 4 \pi v \xi/3$ and $v$ is the Fermi velocity at  
the points where $\epsilon_k = \epsilon_{k+Q} =\mu$. 
For the $\epsilon_k$ which
we are using, these points (hot spots) are 
located near $(0,\pi)$ and symmetry related points. 

The fermionic self-energy in the spin-fermion model is highly 
nontrivial even for  finite $\xi$ due to a hidden singularity at $\omega
\rightarrow 0$ which needs to be regularized, and has the form
\begin{eqnarray}
\Sigma (k,\omega) &=& -\left(\frac{g}{g_0}\right)^2 
[\frac{2 \omega}{1 + \sqrt{1 -i
|\omega|/\omega_{sf}}}~\times \nonumber \\ 
&&\Phi_1 \left(\frac{\epsilon^2_{k+Q} \xi^2 \omega_{sf}}{v^2 |\omega|}\right)
- \epsilon_{k+Q} \Phi_2 \left(\frac{v \xi^{-1}}{\omega_{sf}}\right)]
\label{se}
\end{eqnarray}
where the two scaling functions have the following limiting
behavior: $\Phi_1 (0) =1, ~\Phi(x \gg 1) \sim x^{-1/2},~\Phi_2(0) =1, 
~\Phi_2 (y \geq 1) = 4 \ln y /(\pi y)$. 
Apparently, $\Sigma(k,\omega)$  is large for $g \gg g_0$. 
However, substituting $\omega_{sf}$ into
the scaling functions, we find that for $g \gg g_0$, $y \gg 1$ and hence
$\Phi_2 (y) \ll 1$. In this situation, $\Sigma(k,0)$ depends on $g$ only
logarithmically and in fact saturates if we impose an upper cutoff in
the spin susceptibility at $\omega_{max} \sim 2J$.
 We computed $\Phi_2 (y)$ beyond
logarithmical accuracy and found that $(g/g_0)^2 \Phi_2 (y)$ is always smaller
than $1$ (it saturates at about $0.4$ for $g/g_0 \rightarrow \infty$). 
In this situation, the Fermi surface evolution wouldn't start, and one
preserves a large, Luttinger-type Fermi surface~\cite{morr}.

On the other hand, the 
frequency dependent term in $\Sigma$ 
still scales as $(g/g_0)^2$ in a region where  $x \leq 1$, i.e., where 
$\epsilon^2_{k+Q} \leq 2 v^2 (\omega/\omega_1)$ where $\omega_1 = 
2 \omega_{sf} \xi^2$. In this range, the 
bare $\omega$ term becomes overshadowed by the self-energy 
for $g > g_0$.  Moreover, for $\epsilon^2_k <
(9/8) v^2 (\omega/\omega_1)$, 
the self-energy overshadows both $\omega$ and $\epsilon_k$,
 and to a good accuracy, the renormalized $G$ acquires a universal,
momentum independent form
\begin{equation}
G^{-1}(\omega) = -\Sigma(\omega) = \frac{1}{Z}
~\frac{2\omega}{1 + \sqrt{1 - i|\omega|/\omega_{sf}}}
\label{G}
\end{equation}
where $Z = (g_0/g)^2$. The two conditions on $\epsilon_k$ and $\epsilon_{k+Q}$
select a region around a hot spot with the width
 $\sim \omega/\omega_{1}$. We will see below that the dominant
contribution to the LEG comes from the frequencies $\omega \sim
\omega_1$. For these frequencies, Eq. (\ref{G}) is valid
over a substantial fraction of the Brillouin zone which e.g.
includes the $(0,\pi)$ point.

Eq. (\ref{se}) is obtained to second order in $g$
 but using the renormalized form of
the spin susceptibility. It turns out~\cite{andrey,lev} that
higher-order self-energy and vertex
corrections to both, fermionic self-energy and  spin damping scale 
in the same way as the
momentum dependent term  in (\ref{se}), i.e., 
they depend on $g$ only logarithmically
and in practice reduce to just constants.
 We have checked
that numerically, all third-order corrections are rather small and can be
safely neglected. For example,  for $(g/g_0)^2 =3$, the inclusion of the
vertex correction into the self-energy 
yields only $4\%$ correction to Eq. (\ref{G})~\cite{vertex}.
\begin{figure} [t]
\begin{center} 
\leavevmode
\epsffile{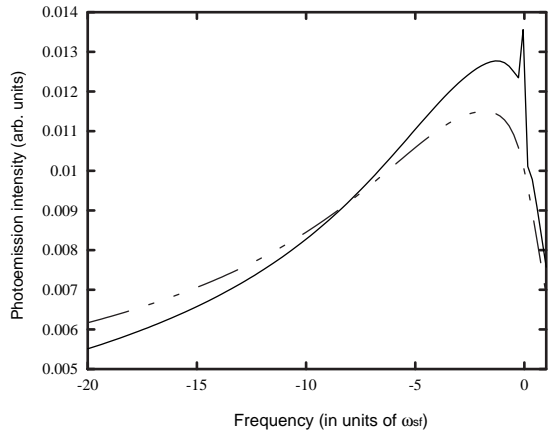}
\end{center}
\caption{The photoemission intensity $I(\omega) = A(\omega) n_F (\omega)$ 
near the Fermi surface for the full solution of the Dyson equation 
with $\Sigma$ given by (\protect\ref{se})
and $(g/g_0)^2 =3$ (solid line), and its strong coupling version given by 
(\protect\ref{G}) (dashed line). We set $T= 2\omega_{sf}$. 
The form of $I(\omega)$ is very similar to the measured photoemission intensity
near $(0,\pi)$ at optimal doping.}
\label{fig1}
\end{figure}
$~~~$We now discuss which ratio $g/g_0$ we expect for cuprates.
For optimally doped $2212Bi$ materials, 
photoemission data imply that $v \approx 1.2
t \sim 0.4 eV$~\cite{flatdisp}.
 Using $\xi \sim 2.5$ inferred from NMR in 214 and 123
 materials near
$T^*$ at optimal doping~\cite{barzykin},
 we obtain $\omega_{sf} \sim 30 (g_0/g)^2 meV$. 
Experimentally, $\omega_{sf} \sim 10 meV$~\cite{barzykin}
 which implies that $(g/g_0)^2 \sim 3$.
Underdoped materials should have even larger
ratio of $g/g_0$. For $(g/g_0)^2 \geq 3$, we found 
that the full $G(k,\omega)$ and
its strong-coupling version, Eq. (\ref{G}), yield virtually 
equivalent results for the spectral function (see Fig.~\ref{fig1}), i.e., 
for this $g/g_0$ the
spectral weight is almost completely transformed into
 the incoherent part of $G$, and the quasiparticle peak is
hardly visible.

Notice that though the 
Green's function in Eq. (\ref{G})
does not give rise to a quasiparticle peak, 
it has a  non-Fermi-liquid form 
$G \propto e^{-i\pi/4} (|\omega| \omega_{sf})^{-1/2} sgn
\omega$ only for $|\omega| > \omega_{sf}$~\cite{comm}.
For smaller $\omega$, we have~ $G^{-1} \propto (\omega + i \omega
|\omega|/(4\omega_{sf}))$. This is a 
conventional Fermi-liquid form of the fermionic 
Green's function right at the Fermi surface. 
The  peculiarity of the present case is
that  the strong self-energy corrections effectively 
freeze  the system at the Fermi surface even if actual $k$ deviates 
from $k_F$ and moves over the region where Eq. (\ref{G}) is valid.
Away from this region,  the self-energy corrections get smaller and one should
recover some renormalized dispersion on a scale of $t$. Notice that
this  behavior is fully consistent
with the ``flat dispersion'' observed near optimal doping~\cite{flatdisp}.

Having obtained the form of the quasiparticle Green's function, we now consider
what happens in the pairing channel. The pairing interaction
is obtained from (\ref{1}) in the same way as in the BCS theory, 
the only difference is that here
the intermediate
boson is a paramagnon rather than a phonon.
It has been several times in the literature that this interaction
is attractive in $d_{x^2 -y^2}$ channel and repulsive 
in all other channels~\cite{d-wave}.
 The
$d-$wave component of the pairing interaction has a form
$\Gamma_{d_{x^2 -y^2}} (k,-k,p,-p,\omega) = d_k~d_p~\Gamma(|\omega|)$ where
$d_k = (\cos k_x - \cos k_y)$ and
$\Gamma(|\omega|)$ is a decaying function of the transferred 
frequency with the limiting
behavior $\Gamma(0) \propto \ln \xi$ and $\Gamma(|\omega|) 
\propto \omega^{-2}$ for $|\omega| \gg \omega_1$.
Numerically, $\Gamma(|\omega|)$ is rather flat for $|\omega| <
\omega_1$ and decreases at higher frequencies. To simplify the analysis, 
we set $\Gamma(|\omega|) = \Gamma = 
const$ for $|\omega| < \omega_1$ and zero for
$|\omega| > \omega_1$.
 The constant is chosen such that the area under $\Gamma(\omega)$
is the same as in the exact expression. This procedure yields 
$\Gamma = -0.16 (g/\xi)^2 = - 3.57~\omega_{sf}/Z^2$. Using this approximation, we explicitly 
can  sum up RPA series in the
particle-particle channel and obtain a $d_{x^2 -y^2}$ pairing susceptibility   
in the form $\chi^{sc} (-k +q,k,\Omega) = d^2_k~\chi^{sc}
(q,\Omega)$ where
\begin{equation}
\chi^{sc} (q,\Omega) = \frac{3}{2}~\Gamma \frac{
3 \Gamma
\Pi(q,\Omega)}{(1+3\Gamma \Pi(q,\Omega))}
\label{chi}
\end{equation}
and
$\Pi (q,\Omega) = \int^{\prime} ~d^2_p~
G(\omega) G(\Omega -\omega) ~ d {\vec p} d \omega$
 is a $d-$wave 
polarization operator. The prime to the integration sign indicates
 that the momentum integration goes over the
region where Eq. (\ref{G}) is valid. In this region, the polarization operator
is independent on $q$.  

We now compute $\Pi(0,0)$
and show that for the fully incoherent $G$ from (\ref{G}),
 the $d-$wave pairing interaction 
is strongly enhanced, but there is no real instability upto $T=0$. 
Indeed, 
the momentum integration in the polarization operator goes over the area
$ \sim \omega/\omega_1$. Integrating then over frequency 
we obtain $\Pi \sim Z^2/\omega_{sf} 
\int_{\omega_{sf}}^{\omega_1} (d\omega/\omega) (\omega/\omega_1)$. This integral is clearly
 dominated by $\omega \sim \omega_1$, and yields
 $\Pi \sim Z^2/\omega_{sf} \sim \Gamma^{-1}$, 
i.e., $\Gamma \Pi (0,0) = O(1)$ independent on $g$.
Collecting all numbers, we obtain $3\Gamma \Pi(0,0) \approx -0.7$.
We see that $1+3\Gamma \Pi(0,0)$ is reduced but still remains positive, 
i.e., fully
incoherent $G$ does not give rise to actual superconductivity. 
It does however give rise to  $d-$wave precursors as we now show.  
For this, we construct the 
pairing self-energy using $\chi^{sc}$ and obtain
\begin{equation}
{\bar G}^{-1} (k,\omega) = G^{-1}(\omega) + 
d^2_k~\int \chi^{sc} (q,\Omega) 
G(-\omega +\Omega)
\label{sc}
\end{equation}
Here ${\bar G}$ is the full quasiparticle Green's function, and $G$
plays the role of the bare Green's
function for the Cooper channel.
For the 
$\delta-$functional form of $\chi^{sc} (q,\Omega)$, 
Eq. (\ref{sc}) reduces to a
conventional Gorkov's equation for the full ${\bar G}$.
 In our case, $\chi^{sc}$ is
enhanced, but it 
never acquires a $\delta-$functional peak. Nevertheless, we
can do the same 
trick as with the SDW precursors~\cite{morr}: expand $G(-\omega + \Omega)=
G(-\omega) + G^{\prime}$ and check whether $G^{\prime}$ is relevant.
Without $G^{\prime}$, (\ref{sc}) has the same form as 
in the true superconducting state.
The relative corrections due to $G^{\prime}$ depend the
ratio of the typical width of $\chi^{sc} (q,\Omega)$ and the
typical frequency shift obtained by solving the Gorkov's equation without
$G^{\prime}$. If this ratio is small, then the corrections
are also small. Physically, this means that when the amount of a shift
is larger than the width of a pairing susceptibility, the latter can be
approximately considered as a $\delta-$function at the energy scales
comparable to the shift. 
\begin{figure} [t]
\begin{center}
\leavevmode
\epsffile{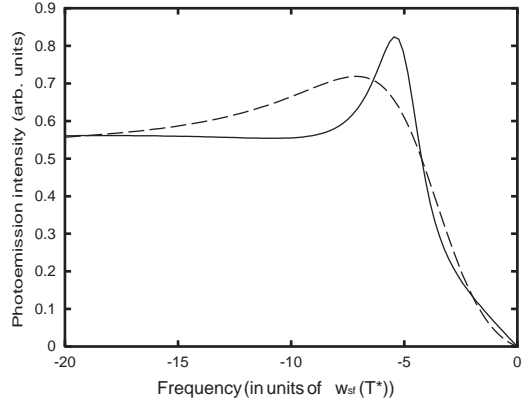}
\end{center}
\caption{The photoemission intensity  obtained from 
(\protect\ref{res}) for $b=3$ (dashed line)
and $b=0.5$ (solid line). The solution for $b=3$
is valid for $T > T_c$ and 
possesses  a LEG and a broad maximum at larger frequencies. 
The solution for $b=0.5$ is
valid for $T\ll T_c$ and possesses a quasiparticle peak. 
As $T$ goes below $T_c$, one
solution gradually moves into the other.}
\label{fig2}
\end{figure}

We now present the results of computations. Let 
us first neglect $G^{\prime}$.
Solving (\ref{sc}) we then obtain
\begin{equation}
{\bar G} ({\bar \omega}) \propto \frac{{\bar\omega}~ 
 (1 +\sqrt{1 - i |{\bar \omega}|})}{{\bar \omega}^2 -
 b^2_k (1 +\sqrt{1 - i |{\bar \omega}|})^2}
\label{res}
\end{equation}
 where ${\bar
\omega} = \omega/\omega_{sf}$ and $b_k  =Z \Delta_d (k)/2 \omega_{sf}$
where $\Delta^2_d (k) = d^2_k~\int d{\vec q} d \Omega ~\chi^{sc} (q,\Omega)$.
The $q-$integration in the last formula again
runs over the area where Eq. (\ref{G}) is valid and 
for $\Omega \sim \omega_1$ 
which, as we will see, dominates the frequency integral, yields $O(1)$.
The integration over $\Omega$ is not formally restricted, 
but the polarization operator
decreases with increasing $\Omega$ such that one
progressively looses the enhancement in the $d-$wave channel. We found that
the polarization operator changes sign at $|\Omega| = \omega_1$.
 To a reasonable accuracy, we can then approximate $\Pi (\Omega)$ as $\Pi (\Omega) = 
\Pi(0)(1 - (\Omega/\omega_1)^2)$. Substituting this result into 
$\chi^{sc}$, performing the integration and collecting all numbers,
 we obtain $b_k  = 0.82 \xi d_k$ which is a large number
near $(0,\pi)$ (notice that $b_k$ does not depend on $g$). 
A simple analysis then shows that the pole in
${\bar G}$ is located almost along imaginary frequency axis, at
$\omega = - i \omega^* (k)$, where $\omega^* (k) = \omega_{sf} b^2_k =
0.34 d^2_k~\omega_1 $.
 As a result, the spectral function
which emerges from (\ref{res}) does not acquire a quasiparticle pole but 
rather a shift by $\omega \sim \omega^*$. For $k$ near $(0,\pi)$ we then have
\begin{equation}
A (\omega) \propto \sqrt{\omega}~\frac{\omega + \omega^* (k)}{\omega^2 +
(\omega^* (k))^2}.
\label{res2}
\end{equation} 
Eqs. (\ref{res}) and (\ref{res2}) are the key results of the paper.
We see that already in the normal state 
the spectral function  rapidly (as  $\sqrt{\omega}$)
increases at low frequencies, reaches half a maximum at $\omega \approx 0.2
\omega^*$, then passes through a maximum at $\omega = \omega^*$, and very 
slowly decreases reaching half a maximum only at $\omega \sim 5 \omega^*$.
 This behavior has a striking resemblance with 
the LEG behavior observed in photoemission (see Fig.~\ref{fig2}).
The position of LEG coincides with the half-maximum at low
frequencies; the broad maximum is located at frequencies which are 
few times larger. This is quite consistent with the data. 
The magnitude of the LEG,
$\omega^{le} \approx 0.54 \omega^{sf} \xi^2$ also has the same order
 of few tens of meV as in the data.

We now estimate the corrections due to $G^{\prime}$. Formally,
the frequency shift and the width 
of $\chi^{sc}$ are both of the order of $\omega_1$. However, 
if the pairing susceptibility is strongly enhanced such that
$1+3\Gamma \Pi(0) = \delta \ll 1$, then the width of the pairing susceptibility
scales as $\omega_1 \sqrt{\delta}$. In this situation,
the relative corrections due to $G^{\prime}$ 
scale as $\delta^{1/2}$ and are small. In our case, $\delta \approx 0.3$. We
 computed the leading
correction to the pairing self-energy due to $G^{\prime}$ and found that 
near $(0,\pi)$ it accounts for $\sim 50\%$ correction for 
$\omega = \omega^{le}$,
and for only $10\%$ correction for $\omega = \omega^*$. 
Though corrections are not that
small, we expect that they somewhat reduce  the amplitude of the LEG, but
do not change substantially the overall shape of $A(\omega)$. 

So far we completely neglected the coherent part of $G$, $G_{coh} =
Z/(i\omega - Z \epsilon_k)$.  This piece contributes a conventional,
logarithmical in $T$ term
to the polarization operator and therefore
  gives rise to a finite $T_c$.
We computed $T_c$ in a standard manner and found
$T_c \sim v Z e^{-c\xi}$ where $c \approx 1$. We see that as the 
doping decreases, $T_c$ actually goes down because the
correlation length increases. Suppose now we are  below $T_c$. Then
the opening of the superconducting gap yields 
a strong negative feedback effect on the spin
damping. This 
gives rise to a rapid increase of $\omega_{sf}$ compared to the perturbative
result, and, hence, to a decrease in 
$b_k$.  The latter yields a gradual shift of the
 pole in ${\bar G}$ in (\ref{res}) from an imaginary to a real axis,
which gives rise to a  gradual transformation of the LEG into
the quasiparticle peak (see Fig. ~\ref{fig2}). This is precisely what has been
observed in the experiments~\cite{camp,shen}.
At $T \ll T_c$, $b_k \ll 1$, and 
 the typical frequencies for the pairing problem are
much smaller than $\omega_{sf}$. At these frequencies, we have a conventional
attractive Fermi liquid, which is just frozen at the Fermi surface in some
$k-$range.  The pairing then gives rise to a conventional quasiparticle pole
at $\omega = \omega^{qp} = 2 b_k \omega_{sf} =
Z \Delta^d (k)$ (we assume that the total spectral weight in $\chi^{sc}$ does
not change as $T$ goes below $T_c$).
Notice that the position of the pole does not depend 
on $\omega_{sf}$ and hence  does not change with temperature. 
We therefore can directly 
compare the 
locations of $\omega^{qp}$ and $\omega^{le}$. 
Substituting the numbers, we find
$\omega^{qp} = (6/\xi (T_c)) \omega^{le}$.
 For underdoped cuprates, $\xi (T_c) \sim 4-5$. 
Then $\omega^{qp}$ is slightly larger than $\omega^{le}$ which fully
agrees with the data.

To summarize, in this paper we have shown that the exchange of 
magnetic fluctuations can account
for the observed LEG in underdoped cuprates. The LEG  
and the broad
maximum of the spectral function at $\sim 5$ times 
larger frequencies turn out to be
byproducts of the same physical effect. 

A final point. In the above discussion 
we assumed that the Fermi surface is not modified by interactions.
In fact, when LEG is formed, the 
spin damping goes down because of a feedback effect from the LEG,
 and the momentum-dependent piece of the self-energy  
increases. 
Eventually, this increase  should trigger
 the evolution of the Fermi surface towards small hole pockets. This evolution
is accompanied by the
suppression of the $d-$wave attraction by 
vertex corrections~\cite{morr,schrieffer}. This last effect
is probably relevant only at very low densities, but it nevertheless
suppresses not only  $T_c$ but also  the LEG before the system reaches
half-filling. How precisely this happens, however, requires further study.

It is my pleasure to thank G. Blumberg, J.C. Campuzano, S. Chakravarty,
A. Finkelstein, R. Joynt, R. Laughlin, Yu. Lu, A. Millis, D. Morr, M. Norman, 
M. Onellion, D. Pines, S. Sachdev, J. Schmalian, Z-X Shen, Q. Si,
 B. Stoikovic and A. Tsvelik
 for useful conversations. The research was supported by NSF DMR-9629839.
The author is an A.P. Sloan fellow.

\end{document}